\documentclass[pra,twocolumn,showpacs]{revtex4}
\usepackage{amsmath}
\usepackage{bbm}

\begin{document}

\title{Local phase damping of single qubits sets an upper bound on the phase damping rate of entangled states}
\author{Stephan D\"{u}rr}
\affiliation{Max-Planck-Institut f{\"u}r Quantenoptik, Hans-Kopfermann-Stra{\ss}e 1, 85748 Garching, Germany}

\pacs{03.65.Yz, 03.65.Ud}

\begin{abstract}
I derive an inequality in which the phase damping rates of single qubits set an upper bound for the phase damping rate of entangled states of many qubits. The derivation is based on two assumptions, first, that the phase damping can be described by a dissipator in Lindblad form and, second, that the phase damping preserves the population of qubit states in a given basis.
\end{abstract}

\maketitle

\section{Introduction}

Quantum information processing \cite{nielsen:00} offers a perspective for a tremendous reduction of the computation time in solving certain problems, such as factorization of large numbers \cite{shor:94}, simulations of quantum systems \cite{feynman:82}, and database searches \cite{grover:97}. However, interactions of the quantum system with its environment induce decoherence \cite{zurek:91, haroche:98, zurek:03, zurek:09}. This is a major limiting factor on the way toward large-scale experimental implementations. Naively, one might expect that the decoherence rate of an entangled many-qubit state should equal the sum of the decoherence rates of the individual qubits. While this is true if the decoherence processes are local, i.e.\ if they act independently on individual qubits, the situation can change if the decoherence processes act on some or all qubits in a correlated manner. Such correlated decoherence can give rise to decoherence-free subspaces which have been studied theoretically \cite{palma:96, zanardi:97, duan:97, lidar:98, blume-kohout:08} and experimentally \cite{kwiat:00, kielpinski:01, fortunato:02, carravetta:04}. This shows that certain entangled states can decohere considerably slower than their single-qubit constituents. In other words, a single-qubit decoherence rate is not a \emph{lower} bound for the decoherence rate of entangled states.

Here I show that the single-qubit decoherence rates set a rigorous \emph{upper} bound for the decoherence rate of entangled states of $n$ qubits. The existence of such an upper bound is somewhat surprising because entanglement is difficult to prepare and maintain experimentally. The upper bound derived here is experimentally relevant because in many experiments where one aims at generating entangled states one has the capability to measure single-qubit decoherence rates already during the build-up phase of the apparatus, often much earlier than the time where one manages to generate and detect entangled states. In addition, one often uses experimental techniques such as a magnetic hold field to suppress spin-flip transitions of the qubits. But these techniques cannot protect the qubit against loss of phase coherence. The experimenter thus selects a preferred basis in which the populations are preserved, whereas phase coherence may decay.

\section{Model}

The derivation of the upper bound rests on two assumptions. The first assumption is that there is an orthonormal basis $\mathcal B$ in which the population of the basis states is time independent whereas the relative phases between the basis states may rotate and decay. This condition is easy to verify experimentally. I will show below that this requires the Hamiltonian $H$ to be diagonal in the basis $\mathcal B$. This implies that the final results will be relevant for quantum memories, but they are not necessarily applicable during quantum gate operations that act on the qubits of interest. The second assumption is that the decoherence can be described by a dissipator in Lindblad form. This is a reasonable assumption for many experiments, as discussed now.

Consider a quantum system coupled to an environment. If one assumes, first, that the system is initially not entangled with the environment and, second, that a Born-Markov approximation is appropriate (because the coupling between system and environment is weak enough and because the environment is much larger than the system), then the environmentally-induced decoherence can be described by
\begin{subequations}
\label{Lindblad}
\begin{eqnarray}
\frac{d}{dt} \rho
= \frac1{i\hbar} [H,\rho] + \mathcal D \rho
\end{eqnarray}
with a dissipator $\mathcal D$ in Lindblad form \cite{lindblad:76, carmichael:99, breuer:02}
\begin{eqnarray}
\mathcal D \rho
&=& \sum_m \frac{\gamma^{(m)}}2 \left( 2A^{(m)}\rho (A^{(m)})^\dag
\right. \nonumber \\ && \left.
-(A^{(m)})^\dag A^{(m)} \rho- \rho (A^{(m)})^\dag A^{(m)}\right)
.\end{eqnarray}
\end{subequations}
Here, $\rho$ is the density matrix, the $\gamma^{(m)}>0$ are decoherence rates, and the dimensionless operators $A^{(m)}$ are called Lindblad operators. In the following, it is always assumed that the time evolution is described by Eq.\ \eqref{Lindblad}.

\emph{Definition}: I call a time evolution generated by Eq.\ \eqref{Lindblad} \emph{population preserving} with respect to an orthonormal basis $\mathcal B$ if all diagonal elements of $\rho$ in the basis $\mathcal B$ are time independent for all initial states, i.e. $(d/dt)\rho_{ii}=0$.

\section{Results}

I now formulate three theorems, which are proven in appendix \ref{app-proofs}.

\emph{Theorem 1}: If all the $A^{(m)}$ are diagonal in the same basis $\mathcal B$ with eigenvalues $\lambda_i^{(m)}$, then
\begin{eqnarray}
\label{D-rho-ij}
\mathcal D \rho_{ij}
= \left( i \Delta_{ij} - \Gamma_{ij} \right)\rho_{ij}
\end{eqnarray}
with decay coefficients $\Gamma_{ij}$ and angular frequencies $\Delta_{ij}$ given by
\begin{eqnarray}
\label{Gamma-def}
\Gamma_{ij}
&=& \frac12 \sum_m \gamma^{(m)} \left|\lambda_i^{(m)}-\lambda_j^{(m)}\right|^2
,\\
\Delta_{ij}
&=& \sum_m \gamma^{(m)} {\rm Im}\left(\lambda_i^{(m)} (\lambda_j^{(m)})^*\right)
.\end{eqnarray}
If additionally, $H$ is diagonal in the same basis with eigenvalues $E_i$, then
\begin{eqnarray}
\label{phase-damping}
\frac{d}{dt} \rho_{ij}
= \left( i\omega_{ij} - \Gamma_{ij} \right) \rho_{ij}
\end{eqnarray}
with angular frequencies
\begin{eqnarray}
\omega_{ij}
&=& \frac{E_j-E_i}{\hbar} + \Delta_{ij}
. \end{eqnarray}
Note that $\Gamma_{ij}\geq0$ and $\Gamma_{ii}=\Delta_{ii}=\omega_{ii}=0$. Hence, only off-diagonal elements of $\rho$ change over time. These elements do not mix. Instead, each off-diagonal element experiences two effects. The first effect, described by $\Gamma_{ij}$, is an exponential decay of $|\rho_{ij}|$, which is called phase damping \cite{nielsen:00}. The second effect, described by $\omega_{ij}$, is a phase rotation \cite{spin-echo}. This paper focuses on phase damping.

\emph{Theorem 2}: A time evolution is population preserving with respect to $\mathcal B$, if and only if $H$ and all the $A^{(m)}$ are diagonal in $\mathcal B$.

\emph{Theorem 3}: If a time evolution is population preserving with respect to $\mathcal B$, then the $\Gamma_{ij}$ from theorem 1 obey the following inequality for $n\in\{1,2,3,\dots\}$ and for all basis states $|p_0\rangle\in \mathcal B$, $|p_1\rangle\in \mathcal B$, \dots, $|p_n\rangle\in \mathcal B$
\begin{eqnarray}
\label{the-inequality}
\Gamma_{p_0,p_n} \leq n \sum_{k=1}^n \Gamma_{p_{k-1},p_k}
. \end{eqnarray}
This inequality becomes an equality, if and only if for each $m$, the expression $\lambda_{p_{k-1}}^{(m)} - \lambda_{p_k}^{(m)}$ is independent of $k$, where the $\lambda_i^{(m)}$ are the eigenvalues of the $A^{(m)}$.

This theorem is applicable to the density matrix $\rho$ of an arbitrary state of $n$ qubits. Let the states $\protect{|\!\uparrow\rangle}$ and $\protect{|\!\downarrow\rangle}$ denote a basis of the Hilbert space of each single qubit and assume that the time evolution is population preserving with respect to the basis $\mathcal B$ which consists of all the tensor products of the single-qubit states $\protect{|\!\uparrow\rangle}$ and $\protect{|\!\downarrow\rangle}$. This situation corresponds, \emph{e.g.}, to the scenario with a magnetic hold field discussed in the introduction. According to theorem 1, each density matrix element $\rho_{ij}$ in the preferred basis $\mathcal B$ obeys Eq.\ \eqref{phase-damping}. Specifically, there will be phase damping, described by $\Gamma_{ij}$. For any choice of the basis states $|i\rangle$ and $|j\rangle$ one can obviously construct a sequence of basis states $|p_0\rangle$, $|p_1\rangle$, \dots, $|p_n\rangle$ in a way that $|p_0\rangle=|i\rangle$, $|p_n\rangle=|j\rangle$, and that (for all $k\in\{1,2,\dots,n\}$) the states $|p_{k-1}\rangle$ and $|p_k\rangle$ differ by only one local spin flip between states $\protect{|\!\uparrow\rangle}$ and $\protect{|\!\downarrow\rangle}$. Application of Eq.\ \eqref{the-inequality} yields an upper bound for $\Gamma_{ij}$ and each $\Gamma_{p_{k-1},p_k}$ on the righthand side obviously describes a local phase damping rate.

This is the central result of the present paper. Interestingly, the upper bound in the inequality \eqref{the-inequality} is a factor of $n$ higher than the decoherence rate that one would obtain for uncorrelated phase damping of individual qubits.

The experimental relevance of this upper bound arises from the fact that measurements of the single-qubit phase damping rates are fairly easy to perform because no entangled state needs to be prepared and detected. According to the inequality \eqref{the-inequality}, such measurements already set a worst-case upper bound for the phase damping rate of any entangled state.

\section{Examples}

To illustrate this concept, consider an example of a Greenberger-Horne-Zeilinger(GHZ)-type entangled state \cite{greenberger:90} of $n$ qubits
\begin{eqnarray}
\label{GHZ}
|\psi_\mathrm{GHZ}\rangle =
\frac1{\sqrt2} \left( |\!\uparrow\rangle^{\otimes n} + |\!\downarrow\rangle^{\otimes n} \right)
. \end{eqnarray}
Here, $\protect{|\!\uparrow\rangle}^{\otimes n}=\protect{|\!\uparrow\rangle} \otimes \cdots \otimes \protect{|\!\uparrow\rangle}$ abbreviates a tensor product of $n$ times the same single-qubit state and the inequality \eqref{the-inequality} can be applied to the basis states
\begin{eqnarray}
|p_k\rangle = |\!\uparrow\rangle^{\otimes k} \otimes |\!\downarrow\rangle^{\otimes (n-k)}
\end{eqnarray}
with $k\in\{0,1,\dots,n\}$. This yields an upper bound for the phase damping rate of the GHZ state because $\Gamma_{p_0,p_n}$ describes the decay of
$\langle p_0|\rho|p_n\rangle=\langle \downarrow\cdots\downarrow|\rho |\uparrow\cdots\uparrow\rangle$. The other coefficients $\Gamma_{p_{k-1},p_k}$ appearing in the inequality \eqref{the-inequality} describe decay of $\langle p_{k-1}|\rho|p_k\rangle$. The states $|p_{k-1}\rangle$ and $|p_k\rangle$ differ only by a spin flip of the $k$-th qubit. Hence, this coefficient can be determined experimentally from a measurement of the single-qubit phase damping rate of the $k$-th qubit.

Let $\Gamma_\mathrm{GHZ}=\Gamma_{p_0,p_n}$ denote the phase damping rate coefficient of the GHZ state and let $\Gamma_k=\Gamma_{p_{k-1},p_k}$ denote the phase damping rate coefficient of the $k$-th qubit, where the specific orientation of the other qubits was dropped from the notation for brevity. Then Eq.\ \eqref{the-inequality} yields
\begin{eqnarray}
\label{GHZ-inequality}
\Gamma_\mathrm{GHZ}\leq n (\Gamma_1+\Gamma_2+\dots+\Gamma_n)
.\end{eqnarray}

To illustrate this inequality further, consider some examples for $n=2$ qubits. Here, one typically denotes the Bell states as
\begin{eqnarray}
|\psi^\pm \rangle &=& \frac1{\sqrt2}(|\!\uparrow\downarrow\rangle\pm|\!\downarrow\uparrow\rangle)
,\\
|\phi^\pm \rangle &=& \frac1{\sqrt2}(|\!\uparrow\uparrow\rangle\pm|\!\downarrow\downarrow\rangle)
.\end{eqnarray}
For 2 qubits, the GHZ-type state of Eq.\ \eqref{GHZ} is obviously the Bell state $|\phi^+\rangle$.

\emph{Example 1}: Consider local phase damping generated by the Lindblad operators
\begin{eqnarray}
A^{(1)}
&=& |\!\uparrow\rangle\langle\uparrow\!|_1 \otimes \mathbbm 1_2
,\\
A^{(2)}
&=& \mathbbm 1_1 \otimes |\!\uparrow\rangle\langle\uparrow\!|_2
\end{eqnarray}
where $\mathbbm 1$ denotes the identity matrix. This yields $\Gamma_1=\gamma^{(1)}$, $\Gamma_2=\gamma^{(2)}$, and
\begin{eqnarray}
\Gamma_{\psi^\pm}=\Gamma_{\phi^\pm}=\Gamma_1+\Gamma_2
.\end{eqnarray}
Local phase damping acts identically on all Bell states. The phase damping rate of each Bell state is simply the sum of the local phase damping rates.

\emph{Example 2}: Consider phase damping generated by the Lindblad operator
\begin{eqnarray}
\label{example-2}
A^{(1)} = |\!\uparrow\uparrow\rangle\langle\uparrow\uparrow\!|
- |\!\downarrow\downarrow\rangle\langle\downarrow\downarrow\!|
.\end{eqnarray}
The states $\protect{|\!\uparrow\uparrow\rangle}$ and $\protect{|\!\downarrow\downarrow\rangle}$ experience phase damping relative to the two-dimensional rest of the Hilbert space. Here, $\Gamma_1=\Gamma_2=\gamma^{(1)}$ and
\begin{eqnarray}
\Gamma_{\psi^\pm}=0
,\qquad
\Gamma_{\phi^\pm} = 2(\Gamma_1+\Gamma_2)
.\end{eqnarray}
The states $|\psi^\pm\rangle$ span a decoherence-free subspace, whereas the states $|\phi^\pm\rangle$ experience phase damping at a rate that reaches the upper bound \eqref{GHZ-inequality} for $n=2$.

This example is somewhat related to the experiment with two trapped ions in Ref.\ \cite{kielpinski:01}. In that experiment, the phase damping is dominated by fluctuating ambient magnetic fields with frequencies primarily at 60 Hz and its harmonics. These fields cause a fluctuating Zeeman energy for the ions. The ions are separated by only a few micrometers so that the ambient magnetic fields are roughly uniform across the trapping region. As a result, the states $\protect{|\!\uparrow\downarrow\rangle}$ and $\protect{|\!\downarrow\uparrow\rangle}$ experience no net Zeeman effect and the states $|\psi^\pm\rangle$ experience no phase damping, at least to lowest order. The states $\protect{|\!\uparrow\uparrow\rangle}$ and $\protect{|\!\downarrow\downarrow\rangle}$, however, experience plus or minus twice the single-qubit Zeeman shift, leading to a phase damping that is twice as fast as for a single qubit.

Note that technical fluctuations of a macroscopic magnetic field need not necessarily allow for a description in terms of Eq.\ \eqref{Lindblad}. However, if a similar experiment were performed with quantum dots in a solid, then the magnetic dipole-dipole interactions with the large number of surrounding nuclear spins with thermal occupation may create an environment that can be described by Eq.\ \eqref{Lindblad}.

\emph{Example 3}: For comparison, consider phase damping generated by the Lindblad operators
\begin{eqnarray}
\label{example-3}
A^{(1)} = |\!\uparrow\uparrow\rangle\langle\uparrow\uparrow\!|
,\qquad
A^{(2)} = |\!\downarrow\downarrow\rangle\langle\downarrow\downarrow\!|
\end{eqnarray}
with $\gamma^{(1)}=\gamma^{(2)}$. Here, $A^{(1)}$ creates phase damping for the state $\protect{|\!\uparrow\uparrow\rangle}$ relative to the three-dimensional rest of the Hilbert space and $A^{(2)}$ creates an analogous effect for the state $\protect{|\!\downarrow\downarrow\rangle}$. This yields $\Gamma_1=\Gamma_2=\gamma^{(1)}$ and
\begin{eqnarray}
\Gamma_{\psi^\pm}=0
,\qquad
\Gamma_{\phi^\pm} = \Gamma_1+\Gamma_2
.\end{eqnarray}
Again, the states $|\psi^\pm\rangle$ span a decoherence-free subspace. But now, the phase damping rate of the states $|\phi^\pm\rangle$ no longer reaches the upper bound \eqref{GHZ-inequality}.

Such a phase damping might be obtained hypothetically, if there are two different fields, one coupling only to state $\protect{|\!\uparrow\uparrow\rangle}$ and another coupling only to state $\protect{|\!\downarrow\downarrow\rangle}$. If these fields are uncorrelated, they will each generate an individual Lindblad operator, as in Eq.\ \eqref{example-3}. This situation differs from the phase damping caused by a common field, which is expressed by one Lindblad operator as in Eq.\ \eqref{example-2}.

\acknowledgments

I would like to thank Simon Baur, Geza Giedke, Markus Heyl, and Stefan Kehrein for stimulating discussions. This work was supported by the German Excellence Initiative through the Nanosystems Initiative Munich and by the Deutsche Forschungsgemeinschaft through SFB 631.

\appendix

\section{Proofs}
\label{app-proofs}

\emph{Proof of theorem 1}: $A^{(m)}_{ij}=\lambda_i^{(m)}\delta_{ij}$ and Eq.\ \eqref{Lindblad} yield $\mathcal D\rho_{ij} = \sum_{klm} \frac{\gamma^{(m)}}2 \delta_{ik} \rho_{kl} \delta_{lj} (2\lambda_i^{(m)} (\lambda_j^{(m)})^* - |\lambda_i^{(m)}|^2 - |\lambda_j^{(m)}|^2) $, where $\delta_{ij}$ denotes the Kronecker symbol. Combination with $|\lambda_i^{(m)}-\lambda_j^{(m)}|^2=|\lambda_i^{(m)}|^2 + |\lambda_j^{(m)}|^2 - 2{\rm Re}(\lambda_i^{(m)} (\lambda_j^{(m)})^*)$ yields Eq.\ \eqref{D-rho-ij}. Combination with $(d/dt-\mathcal D)\rho_{ij} =(E_i\rho_{ij}-\rho_{ij}E_j)/i\hbar$ completes the proof.

\emph{Proof of theorem 2}: If $H$ and all the $A^{(m)}$ are diagonal in $\mathcal B$, then theorem 1 yields $\Gamma_{ii}=\omega_{ii}=0$ and $(d/dt)\rho_{ii}=0$ for all $i$.

To prove the other direction, assume that the time evolution is population preserving with respect to $\mathcal B$. I will first show that all the $A^{(m)}$ are diagonal in $\mathcal B$. To this end, it suffices to consider $(d/dt)\rho_{jj}=0$ for the special case where $\rho=|j\rangle\langle j|$ at $t=0$ where $|j\rangle$ is any pure basis state that belongs to $\mathcal B$. Eq.\ \eqref{Lindblad} yields $0=(d/dt)\rho_{jj} =- \sum_m \gamma^{(m)} \sum_{k\neq j} |A_{kj}^{(m)}|^2$. This holds for all $j$. This yields $A_{kj}^{(m)}=0$ for all $j,k,m$ with $j\neq k$. Hence, all the $A^{(m)}$ are diagonal in $\mathcal B$.

It remains to be shown that $H$ is diagonal in $\mathcal B$. To this end, select an arbitrary pair of indices $i,j$ with $i\neq j$. It needs to be shown that $H_{ij}=0$. Here, it suffices to consider only initial density matrices of the form $\rho=|\psi\rangle\langle\psi|$ with $|\psi\rangle =(|i\rangle+e^{i\varphi}|j\rangle)/\sqrt2$ with a real parameter $\varphi$. As seen above, all the $A^{(m)}$ are diagonal in $\mathcal B$ so that theorem 1 yields $\mathcal D\rho_{ii}=0$. The population preserving character of the time evolution then yields $0=i\hbar(d/dt)\rho_{ii} =\sum_{k\in\{i,j\}}(H_{ik}\rho_{ki}-\rho_{ik}H_{ki}) =H_{ij}\rho_{ji}-\mathrm{c.c.} =i\mathrm{Im}(H_{ij}e^{i\varphi})$. This holds for all real $\varphi$, especially for $\varphi=0$ and $\varphi=\pi/2$. Hence $H_{ij}=0$.

\emph{Proof of theorem 3}: Assume that the time evolution is population preserving. According to theorem 2, all the $A^{(m)}$ are diagonal in $\mathcal B$. Hence, theorem 1 is applicable. The Cauchy-Schwarz inequality $|\sum_{k=1}^n x_ky_k^*|^2\leq (\sum_{k=1}^n |x_k|^2)(\sum_{k=1}^n |y_k|^2)$ with $x_k=\lambda_{p_{k-1}}^{(m)}-\lambda_{p_k}^{(m)}$ and $y_k=1$ reads $|\lambda_{p_0}^{(m)}-\lambda_{p_n}^{(m)}|^2 = |\sum_{k=1}^n (\lambda_{p_{k-1}}^{(m)}-\lambda_{p_k}^{(m)})|^2\leq n \sum_{k=1}^n |\lambda_{p_{k-1}}^{(m)}-\lambda_{p_k}^{(m)}|^2$. Multiply this by $\gamma^{(m)}/2$. Then sum over $m$ to obtain the inequality \eqref{the-inequality}. It is well known that the Cauchy-Schwarz inequality becomes an equality, if and only if the vectors $(x_1,\dots,x_n)$ and $(y_1,\dots,y_n)$ are linearly dependent. Here, this is equivalent to $x_1=x_2=\dots=x_n$.


\begin{thebibliography}{10}

\bibitem{nielsen:00}
M.~A. Nielsen and I.~L. Chuang, {\em Quantum Computation and Quantum
  Information} (Cambridge University Press, Cambridge, UK, 2000).

\bibitem{shor:94}
P.~W. Shor, In {\em Proceedings of the 35th Annual Symposium on Foundations of
  Computer Science}, edited by S.~Goldwasser (IEEE Computer Society, Los
  Alamitos, CA, 1994) p. 116.

\bibitem{feynman:82}
R.~Feynman, Int. J. Theor. Phys. {\bf 21}, 467 (1982).

\bibitem{grover:97}
L.~K. Grover, Phys. Rev. Lett. {\bf 79}, 325 (1997).

\bibitem{zurek:91}
W.~H. Zurek, Phys. Today {\bf 44} {\rm (10)}, 36 (1991).

\bibitem{haroche:98}
S.~Haroche, Phys. Today {\bf 51} {\rm (7)}, 36 (1998).

\bibitem{zurek:03}
W.~H. Zurek, Rev. Mod. Phys. {\bf 75}, 715 (2003).

\bibitem{zurek:09}
W.~H. Zurek, Nat. Phys. {\bf 5}, 181 (2009).

\bibitem{palma:96}
G.~M. Palma, K.-A. Suominen, and A.~K. Ekert, Proc. R. Soc. London A {\bf 452}, 567
  (1996).

\bibitem{zanardi:97}
P.~Zanardi and M.~Rasetti, Phys. Rev. Lett. {\bf 79}, 3306 (1997).

\bibitem{duan:97}
L.-M. Duan and G.-C. Guo, Phys. Rev. Lett. {\bf 79}, 1953 (1997).

\bibitem{lidar:98}
D.~A. Lidar, I.~L. Chuang, and K.~B. Whaley, Phys. Rev. Lett. {\bf 81}, 2594
  (1998).

\bibitem{blume-kohout:08}
R.~Blume-Kohout, H.~K. Ng, D.~Poulin, and L.~Viola, Phys. Rev. Lett. {\bf 100},
  030501 (2008).

\bibitem{kwiat:00}
P.~G. Kwiat, A.~J. Berglund, J.~B. Altepeter, and A.~G. White, Science {\bf
  290}, 498 (2000).

\bibitem{kielpinski:01}
D.~Kielpinski, V.~Meyer, M.~A. Rowe, C.~A. Sackett, W.~M. Itano, C.~Monroe, and
  D.~J. Wineland, Science {\bf 291}, 1013 (2001).

\bibitem{fortunato:02}
E.~M. Fortunato, L.~Viola, J.~Hodges, G.~Teklemariam, and D.~G. Cory, New J.
  Phys. {\bf 4}, 5 (2002).

\bibitem{carravetta:04}
M.~Carravetta, O.~G. Johannessen, and M.~H. Levitt, Phys. Rev. Lett. {\bf 92},
  153003 (2004).

\bibitem{lindblad:76}
G.~Lindblad, Commun. Math. Phys. {\bf 48}, 119 (1976).

\bibitem{carmichael:99}
H.~J. Carmichael, {\em Statistical Methods in Quantum Optics 1: Master
  Equations and Fokker-Planck Equations} (Springer, Berlin, 1999).

\bibitem{breuer:02}
H.-P. Breuer and F.~Petruccione, {\em The Theory of Open Quantum Systems}
  (Oxford University Press, Oxford, UK, 2002).

\bibitem{spin-echo}
Note that pure phase rotation can lead to a decay of the expectation value of an observable. As an example, consider a large number of spin-$1/2$ particles, initially aligned to maximize the expectation value of the $x$ component of the total spin $\langle S_x\rangle$. Assume that the time evolution is population preserving with respect to the $z$ components of each individual spin-$1/2$ particle. It is well-known from spin-echo experiments that in this scenario, unitary phase rotation generated by a Hamiltonian may cause a decay of $\langle S_x\rangle$. Obviously, pure phase damping could also cause such a decay of $\langle S_x\rangle$. Hence, phase rotation and phase damping can sometimes generate similar effets, at least for a specific observable. The difference between phase rotation and phase damping becomes clear when reconstructing the full density matrix, because for pure phase rotation all the $|\rho_{ij}|$ are time independent, whereas in the presence of phase damping the $|\rho_{ij}|$ decay.

\bibitem{greenberger:90}
D.~M. Greenberger, M.~A. Horne, A.~Shimony, and A.~Zeilinger, Am. J. Phys. {\bf
  58}, 1131 (1990).

\end{thebibliography}
\end{document}